\documentclass[12pt,a4paper]{article}
\usepackage[utf8]{inputenc}
\usepackage[T1]{fontenc}
\usepackage[english]{babel}
\usepackage{amsmath}
\usepackage{amsfonts}
\usepackage{amssymb}
\usepackage{makeidx}
\usepackage{graphicx}
\usepackage[colorlinks=True]{hyperref}
\usepackage[labelfont=bf]{caption}
\usepackage[width=18.00cm, height=26.00cm,margin=1in]{geometry}
\author{Anupam Puwar$^a$ and Rahul Sundar$^b$\\$^a$ - \url{https://anupam-purwar.github.io/page/}, Delhi, India \\$^b$ - \url{https://github.com/RahulSundar}, Chennai, India }
\title{\textbf{Keyword Augmented Retrieval: Novel framework for Information Retrieval integrated with speech interface.}}
\date{}
\begin{document}
	\maketitle
	\begin{abstract}
Retrieving answers in a quick and low cost manner without hallucinations from a combination of structured and unstructured data using Language models is a major hurdle. This is what prevents employment of Language models in knowledge retrieval automation. This becomes accentuated when one wants to integrate a speech interface on top of a text based knowledge retrieval system. Besides, for  commercial search and chat-bot applications, complete reliance on commercial large language models (LLMs) like GPT 3.5 etc. can be very costly. In the present study, the authors have addressed the aforementioned problem by first developing a keyword based search framework which augments discovery of the context from the document to be provided to the LLM. The keywords in turn are generated by a relatively smaller LLM and cached for comparison with keywords generated by the same smaller LLM against the query raised. This significantly reduces time and cost to find the context within documents. Once the context is set, a larger LLM uses that to provide answers based on a prompt tailored for Q\&A. This research work demonstrates that use of keywords in context identification reduces the overall inference time and cost of information retrieval. Given this reduction in inference time and cost with the keyword augmented retrieval framework, a speech based interface for user input and response readout was integrated. This allowed a seamless interaction with the language model. 
	\end{abstract}

Information retrieval is one of the core subjects of the Information systems field and has been explored by various researchers with the twin objective of effective and efficient retrieval of information from massive collections of data. It has very important applications, including web based search engines, document retrieval systems and recommendation systems. The earliest works in this field around the Boolean model helped establish concepts for information retrieval queries, thus enabling precise searching. With the Vector Space Model being proposed by Salton {\it et al.} \cite{salton1975vector}, this field gained a solid mathematical foundation as documents and queries were represented as high-dimensional vectors, enabling relevance ranking based on cosine similarity. Then came, Term Frequency-Inverse Document Frequency \cite{salton1988term}, shortly called TF-IDF which introduced the technique for term weighting, it proposed balancing importance of terms in a document relative to their frequency across the entire corpus. This was followed by the Probabilistic Information Retrieval Model \cite{fuhr1992probabilistic} in the 1990s, which brought in the concept of uncertainty and document ranking based on likelihood estimates. On the web engine based information retrieval two approaches namely BM25 (Best Matching 25) \cite{robertson2009probabilistic} and PageRank Algorithm \cite{brin1998anatomy}, by Larry Page and Sergey Brin became popular with the latter leading to development of Google. Today, the Page rank algorithm is synonymous with web based search engines, it ranks web pages based on their link structure.
\begin{figure*}[htbp!]
    \centering
    \includegraphics[width = \textwidth]{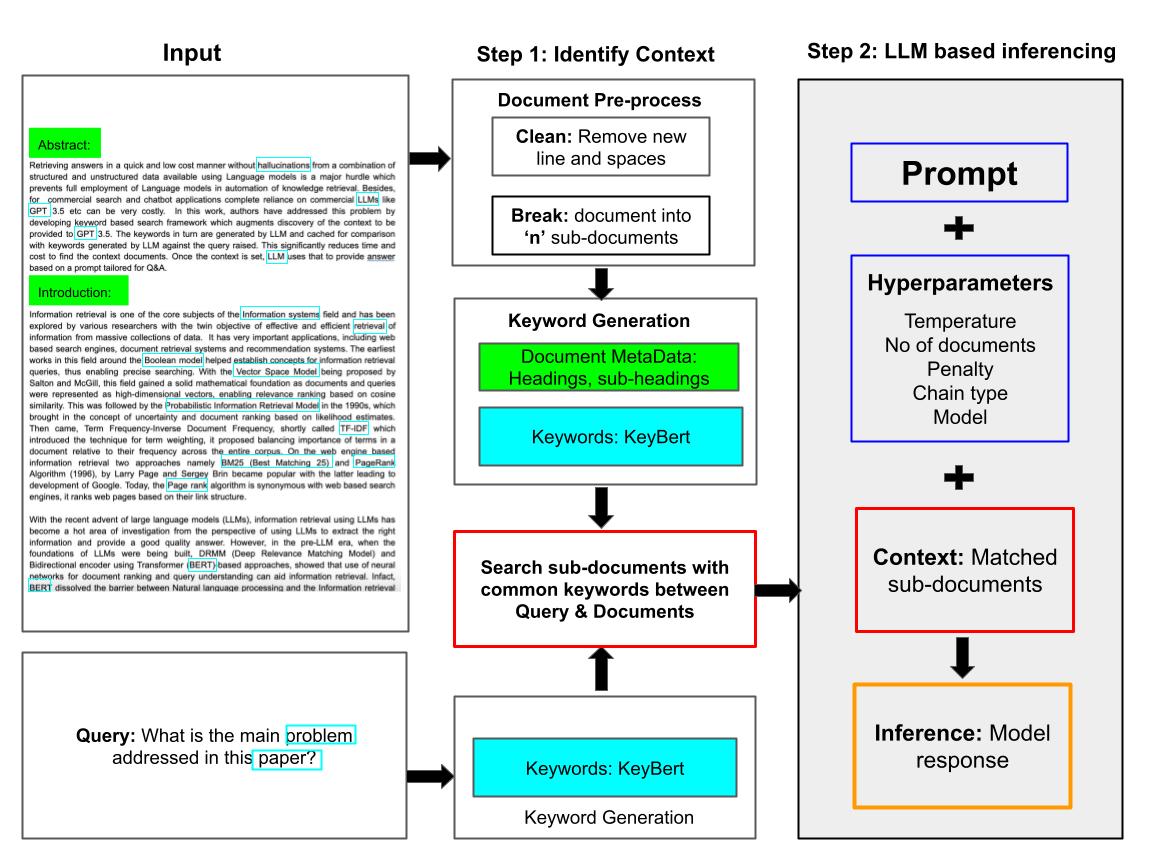}
    \caption{Workflow of KAR}
    \label{fig:enter-label}
\end{figure*}

With the recent advent of large language models (LLMs) \cite{zhao2023survey}, information retrieval using LLMs has become a hot area of investigation from the perspective of using LLMs to extract the right information and provide a good quality answer. However, in the pre-LLM era, when the foundations of LLMs were being built, DRMM (Deep Relevance Matching Model) \cite{Guo_2016} and Bidirectional encoder using Transformer (BERT) \cite{devlin2019bert}-based approaches, showed that use of neural networks for document ranking and query understanding can aid information retrieval. In fact, BERT dissolved the barrier between Natural language processing and the Information retrieval (IR) field. BERT and other transformer based models like GPT~\cite{brown2020language}, 
Google Flan-T5 \cite{weifinetuned} revolutionized the IR field by providing the power of real time interaction to users with their generative and context adapting capabilities.

Commercial LLMs like the ones from OpenAI, Anthropic  have demonstrated promise in terms of providing good answers broadly on many topics. However, even the most sophisticated LLMs have hallucination issues and may end up providing unreasonable answers. Hallucination is the phenomenon where LLM generates false/incorrect or fabricated information. It has been identified as one the primary challenges while using LLM. It puts an unsuspecting user who is depending on these systems for information retrieval, research and critical decision-making at huge risk. Hence, techniques like SelfCheckGPT are being proposed to evaluate the factuality of generated responses~\cite{manakul2023selfcheckgpt} using stochasticity in sampled responses. Besides, LLMs have limited context~\cite{li2023unlocking} and are not aware of latest developments due to the boundary created by the training data they are exposed to. Hence, they sometimes end up cooking up answers which are nothing but manifestations of words with highest probabilities being arranged in an easy to understand language~\cite{manakul2023selfcheckgpt}. Another challenge with the use of LLMs for IR tasks at scale is the high cost associated with commercial LLMs and high inference time/compute cost with open source LLMs. In this perspective, development of a technique which quickly provides a factually correct answer in an affordable manner becomes the need of the hour. In this work, authors propose a novel technique which augments the information retrieval using keywords and headings/labels available within the document.

\section{Methodology}
In the present work, we have devised a two step approach. In step 1, it focuses on understanding the question/user query and generating keywords from it followed by identification of right context from the document search space. Once the right context is identified, an answer is generated by the LLM using the context in step 2. The main contribution of the authors lies in the proposal to identify the right context using keywords generated by a transformer model KeyBert~\cite{grootendorst2020keybert} (an open source Python library) and using these keywords alongwith in-text metadata like section headings to identify the right context from the document space in a quick and low cost manner. 
\subsection{Context identification}
The document database consisting of all documents with text data as well as corresponding meta data is consumed by the model and this massive text data is broken down into smaller sub documents. The length of each sub-document is decided by the context length of the chosen LLM. All these sub-documents are stored and then keywords are extracted from each of them using keyBert library. For each sub-document, the corresponding heading as well as subheading data is also stored as keywords. Thus, it results in a dictionary consisting of each sub- document as index and corresponding keywords. Similar exercise is carried out for the query/question asked using keyBERT library to identify important keywords and query-keyword mapping is generated. Next, we compare the keywords corresponding to the query with keywords corresponding to the sub-documents and this helps identify relevant sub-documents for the question asked. This step replaces the usual step of finding out contextual information from document space using similarity between vector embeddings of query and the document. 

\subsection{LLM based inference}
In the next step, contextual information pulled from the document database is consumed by the LLM along with a prompt to come up with a final answer. The length of answer is decided by the context length of the chosen LLM as well as the number of sub-documents which match the question asked. To minimize the randomness in generation of answer, LLM hyper-parameters viz. temperature is kept as 1e-6 and the LLM used here is text-davinci-003. Following prompt is used to bring out a context informed answer from the LLM.

\noindent\textit{ "Answer the question based on the context below,and if the question can't be answered based on the context, say "I don't know"\\\textbackslash n\textbackslash n Context: {context}\textbackslash n\textbackslash n\\\\ Question: {query}\\\textbackslash n Answer:"}

\subsection{Speech models integration}
\begin{figure}
    \centering
    \includegraphics[width=0.75\textwidth]{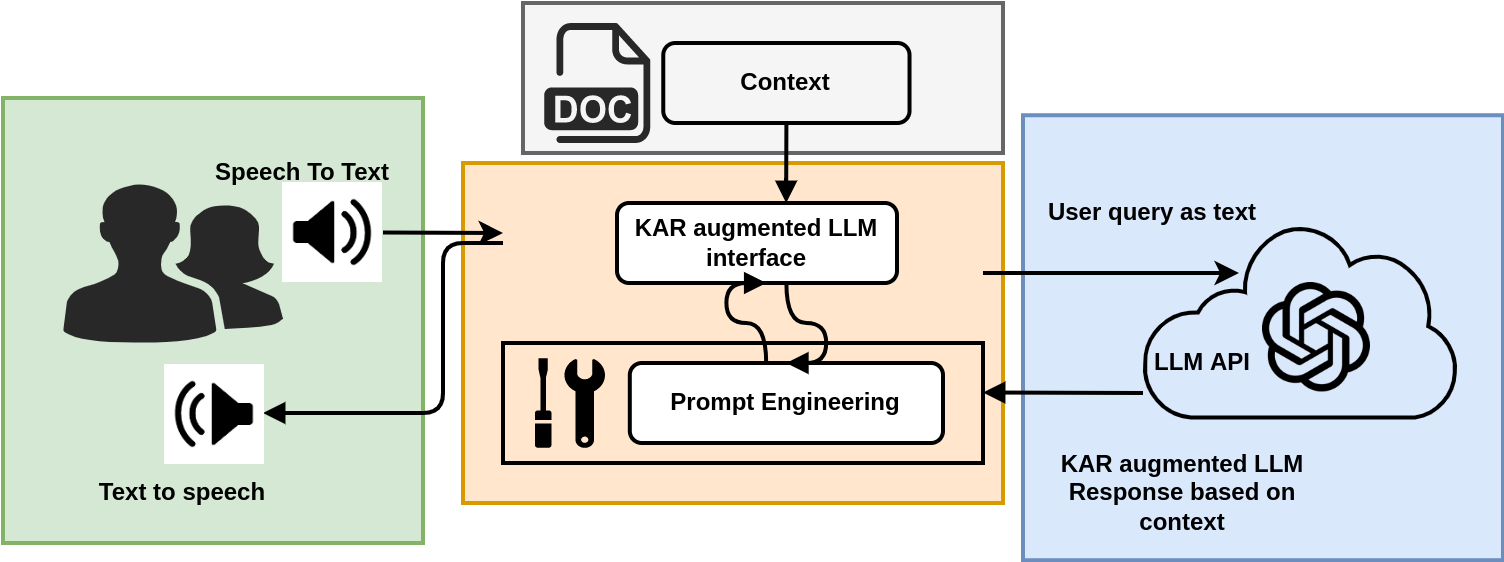}
    \caption{Speech interface integration with the KAR workflow.}
    \label{fig:speechintegration}
\end{figure}

While the KAR is the foundation of the conversational interface, it was investigated how much additional time is added on integration of (i) speech to text models for query input, and (ii) text to speech conversion of the retrieved answer into the system (see figure~\ref{fig:speechintegration}). This is to understand if real time conversational document retrieval is possible through freely available \verb|SpeechRecognition|~\cite{pypiSpeechRecognition} and google text-to-speech (\verb|gTTS|) libraries.   

\section{Results}
The proposed framework, Keyword augmented generation (KAR) consisting of steps $1$ and $2$ was applied to a document (MS word file) created from a web page. The code has been implemented using Python (Langchain~\cite{topsakal2023creating}, keyBERT~\cite{grootendorst2020keybert}, openAI's API~\cite{openaiOpenAIPlatform}, SpeechRecognition~\cite{pypiSpeechRecognition} and gTTS libraries). Inference carried out using context identified using vector embedding similarity between query and document space has been called regular and the one using the proposed Keyword augmented generation framework has been called KAR. Following are the results obtained for the information retrieval task using both approaches, the two primary metrics are inference time and accuracy of answers, as demonstrated by Table 1 below, KAR outperforms on both the metrics. 

Accuracy has been determined based on the completeness of answer from the available information in the source document as well as a binary check on correctness of the answer, if the answer is incorrect, accuracy is marked as 0, if the answer provided by the approach is factually correct and uses the contextual information in the source document but fails to provide an answer based on all the available context, then the accuracy can lie between (0, 100]. 
Here, the KAR approach reduces inference cost as it does not require one to create embeddings for source document and query. Also, one does not need to store source documents as vector databases, however one needs to store keyword-sub-document mapping information in a data-frame to be used during context identification.

\begin{table*}
\caption{Inference time and accuracy metrics for regular and KAR approaches along with inference time for the speech to text and text to speech models.}
\label{tab:inferencetimes}
\resizebox{\textwidth}{!}{\begin{tabular}{p{3cm}p{1.5cm}p{1.5cm}p{1.5cm}p{1.5cm}p{1.5cm}p{1.5cm}p{1.2cm}p{1.5cm}p{1.5cm}}
	\hline
	Query & Answer length (Regular) & Time: Regular (in s) & Answer length (KAR) & Time:KAR (in s) & Time saved & Accuracy (KAR) & Accuracy (Regular) & Time(in s) (STT - Query ) & Time(in s) (TTS - Answer) \\
	\hline
	tell me important facts about expandrank & 305 & 4.034 & 369 & 1.569 & 61.10\% & 100\%
	& 100\%
	& 1.291 & 1.157 \\
	tell me important facts about positionrank & 494 & 3.657 & 592 & 1.682 & 54.00\% & 100\% & 100\%
	& 1.108 & 2.656\\
	what is positionrank & 545 & 4.019 & 331 & 1.696 & 57.80\% & 100\% & 100\%
	& 0.669 & 1.675 \\
	what is page rank & 311 & 2.341 & 282 & 2.005
	& \textbf{14.35}\% & \textbf{85}\% & \textbf{0} & 0.650 & 2.830 \\
	tell me important facts about pagerank
	& 365 & 2.020 & 600 & 2.583
	& \textbf{27.83}\% & \textbf{95}\% & \textbf{0} & 0.8759 & 2.104 \\
	tell me important facts about page rank & 365 & 2.281 & 391
	& 1.352
	& \textbf{40.76}\% & \textbf{75}\% & \textbf{0 }& 1.009 & 1.218 \\
	\hline
\end{tabular}}
\end{table*}

Notably, integration of speech-based query input (STT) and an answer readout stage (TTS) only add approximately 1-2 seconds in inference, respectively as seen in table~\ref{tab:inferencetimes}. However, if the queries were to go beyond a couple of sentences, then the STT response time is expected to go up further. So is the case for answer readout using TTS. Overall, the inference time for the entire speech based-KAR workflow is still of the same order as just regular text-only RAG. 
Cutting down the retrieval time for RAG based systems is thus crucial towards enabling such speech integration for more real-time conversational document retrieval systems. 

\section{Demo}
The KAR framework has been implemented in the chat-bot developed for \href{https://www.aimlsystems.org/2023/}{AIML Systems 2023}. The chat-bot is created using Python and corresponding libraries like Langchain. Its user interface is created and hosted using Streamlit, as shown in Figure~\ref{fig:AIMLSysbot} below.
\begin{figure*}
    \centering
    \includegraphics[width=\textwidth]{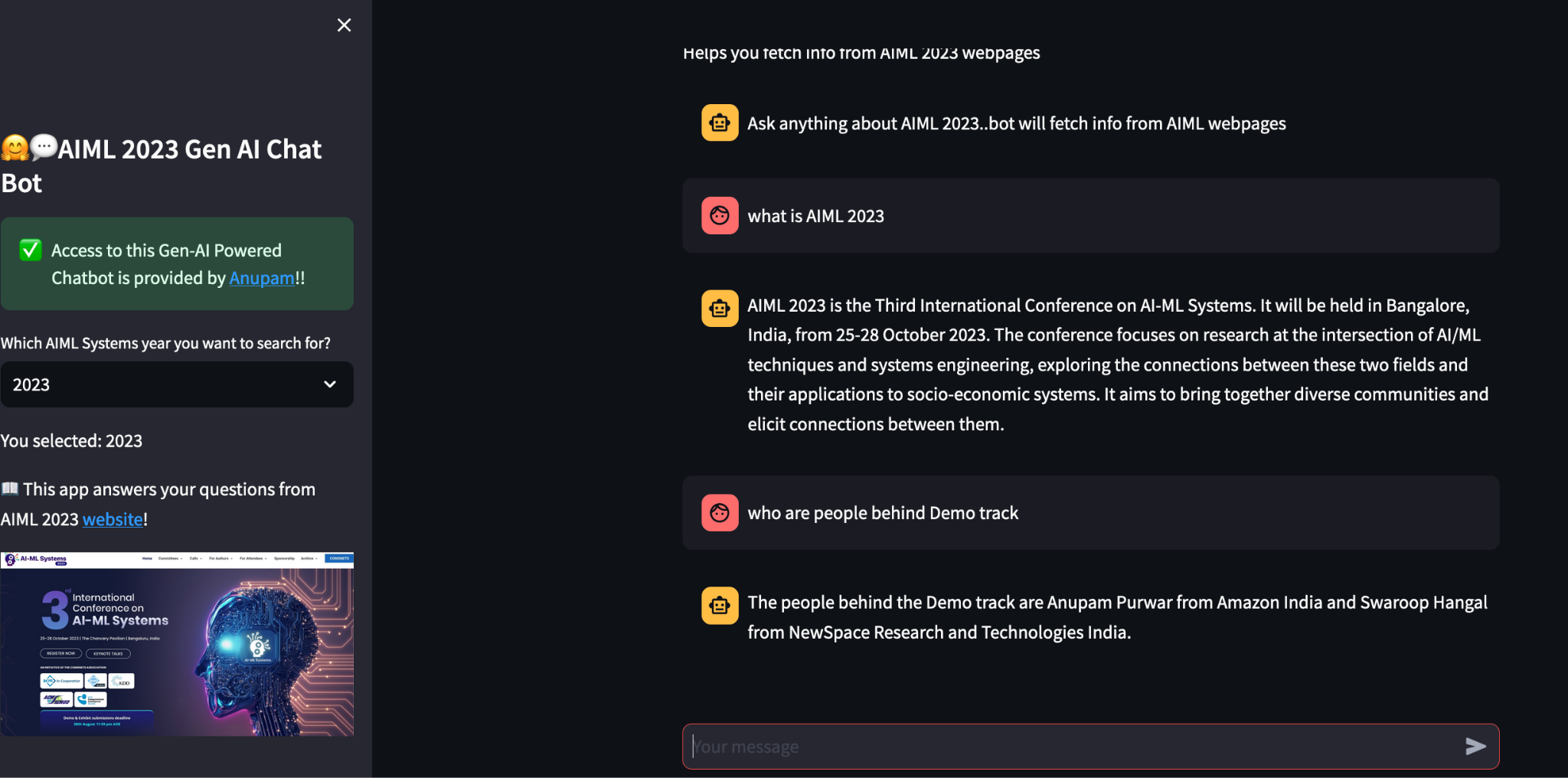}
    \caption{AIML Systems 2023: Generative AI powered bot}
    \label{fig:AIMLSysbot}
\end{figure*}

\section{Conclusions}
This work proposes a novel framework for Information Retrieval from documents using keywords generated by KeyBERT which uses BERT-embeddings as well use of document meta information viz. paragraph headings and subheadings to search and identify the right context corresponding to a user query. Here, a speech based interface was integrated with the keyword augmented retrieval workflow to obtain a more seamless interaction with the user. A speech to text model was used to convert the user speech into a text based query input for the LLM. The KAR augmented LLM responss were then read out using a text to speech model. It was noticed that the speech models along with KAR together would still be comparable to the regular retrieval without speech. The results from implementation of the proposed KAR approach thus  demonstrate better performance in terms of lower inference time, better accuracy and lower inferencing cost. 
In future, authors envisage to extend the framework to temporally resolve the data and make the time context also available to LLM for coming up with accurate answers with time sensitive data as well. 

In addition, the speech interface could also be extended to allow voice cloning and a more human like response with more features including choice between a male and a female voice, and speech translation. 

\bibliographystyle{unsrt}
\bibliography{bibliography}

\end{document}